\documentclass[twocolumn,superscriptaddress,nofootinbib]{revtex4}

\usepackage{graphicx}
\usepackage{dcolumn}
\usepackage{bm}
\usepackage{hyperref}
\usepackage{textcomp}
\usepackage{amsmath}
\usepackage{multirow}

\begin{document}

\title{Spectroscopy of solar neutrinos}

\author{M. Wurm}\email[Corresponding author, e-mail:~]{mwurm@ph.tum.de}
\author{F. von Feilitzsch}
\author{M. G\"oger-Neff}
\affiliation{Physik-Department E15, Technische Universit\"at M\"unchen, James-Franck-Str., D-85748 Garching, Germany}
\author{T. Lachenmaier}
\affiliation{Excellence Cluster Universe, Technische Universit\"at M\"unchen, Boltzmannstr. 2, D-85748 Garching, Germany}
\author{T. Lewke}
\affiliation{Physik-Department E15, Technische Universit\"at M\"unchen, James-Franck-Str., D-85748 Garching, Germany}
\author{Q. Meindl}
\author{R. M\"ollenberg}
\author{L. Oberauer}
\author{W. Potzel}
\author{M. Tippmann}
\affiliation{Physik-Department E15, Technische Universit\"at M\"unchen, James-Franck-Str., D-85748 Garching, Germany}
\author{C. Traunsteiner}
\affiliation{Excellence Cluster Universe, Technische Universit\"at M\"unchen, Boltzmannstr. 2, D-85748 Garching, Germany}
\author{J. Winter}
\affiliation{Physik-Department E15, Technische Universit\"at M\"unchen, James-Franck-Str., D-85748 Garching, Germany}

\date{\today}

\begin{abstract}

In the last years, liquid-scintillator detectors have opened a new window for the observation of low-energetic astrophysical neutrino sources. In 2007, the solar neutrino experiment Borexino began its data-taking in the Gran Sasso underground laboratory. High energy resolution and excellent radioactive background conditions in the detector allow the first-time spectroscopic measurement of solar neutrinos in the sub-MeV energy regime. The experimental results of the Beryllium-7 neutrino flux measurements \cite{arp08b} as well as the prospects for the detection of solar Boron-8, pep and CNO neutrinos are presented in the context of the currently discussed ambiguities in solar metallicity. In addition, the potential of the future SNO+ and LENA experiments for high-precision solar neutrino spectroscopy will be outlined.
\end{abstract}

\maketitle

\section{Introduction}

Solar neutrinos offer a unique way to study the fusion processes in the center of our Sun. The bulk of the released energy, about 98\,\%, is converted into photons, that take about $10^5$ years to reach the solar surface. Contrariwise,  neutrinos released in solar fusion reactions (Sect.\,\ref{SecNeuPro}) leave the Sun unimpeded, taking seconds to arrive at the solar surface and minutes for their travel to the Earth. This free-streaming is based on the fact that neutrinos are only subject to weak nuclear forces, featuring typical interaction cross sections of 10$^{-44}$\,cm$^2$.

The inherently low cross sections pose the main challenge to neutrino detection: In solar neutrino experiments, target masses of several tens or hundreds of tons are required to extract a meaningful signal. A variety of solar neutrino detectors has been devised since the late 1960s (Sect.\,\ref{SecNeuExp}): While in the early stages only integral measurements of the solar neutrino flux were possible, state-of-the-art detectors allow for a time and energy-resolved measurement. The most recent experiment to go into operation is Borexino, a large-volume liquid-scintillator detector allowing the real-time detection of neutrinos in the sub-MeV range (Sect.\,\ref{SecBorex}).

The main concern of solar neutrino detection was for a long time to reveal the particle properties of the neutrino themselves. After the establishment of neutrino oscillations, experiments are now arriving at a stage at which information on the Sun can be extracted from the measured neutrino spectrum. Recently, it has been proposed that neutrinos could be used as indicators of the metallicity of the solar core, as helioseismological and spectroscopic evidence is incongruent (Sect.\,\ref{SecSolMet}). While the available neutrino data is for the moment inconclusive, refinements of the measurements of current experiments might improve the situation. Moreover, future experiments like the SNO+ and LENA detectors will further improve the precision of the solar neutrino measurement. Sect.\,\ref{SecFutExp} will highlight the range of open questions in particle and solar physics that these experiments will be able to answer. 

\section{The Solar Neutrino Spectrum}
\label{SecNeuPro}

The Sun is one of the strongest natural neutrino sources on Earth, producing a flux of 6.6$\times$10$^{10}$ per cm$^{2}$s \cite{bah06}. Neutrinos are generated as a byproduct of the fusion processes fueling the solar energy production, mainly by the proton-proton (pp) fusion chain, but to a small extent (1-2\,\%) also by the catalytic CNO cycle. The net reaction of both processes is
\begin{eqnarray} \label{EqNetRea}
4p \rightarrow {^4\mathrm{He}} + 2e^+ + 2\nu_e + 24.7\,\mathrm{MeV}.\nonumber
\end{eqnarray}
Two neutrinos are generated per completed {$^4$He} fusion.

\begin{figure}[h!]
\begin{centering}
\includegraphics[width=0.47\textwidth]{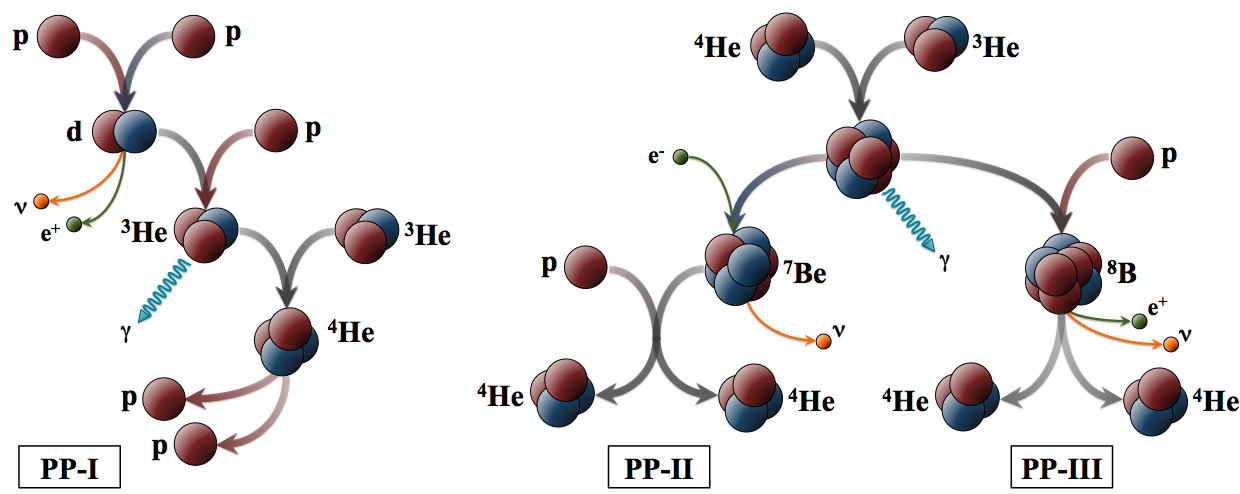}
\caption{The basic reactions of the pp chain: In the pp-I chain, two protons are fused to a deuteron, releasing a positron and a $\nu_e$, the so-called pp neutrino. While the pp-II chain produces a monoenergetic neutrino by the electron-induced conversion of a {$^7$Be} to a {$^7$Li} nucleus, in the pp-III chain high-energetic neutrinos are released by the decay of a {$^8$B} nucleus.}
\label{FigPpChain}
\end{centering}
\end{figure}

\begin{figure}[h!]
\begin{centering}
\includegraphics[width=0.4\textwidth]{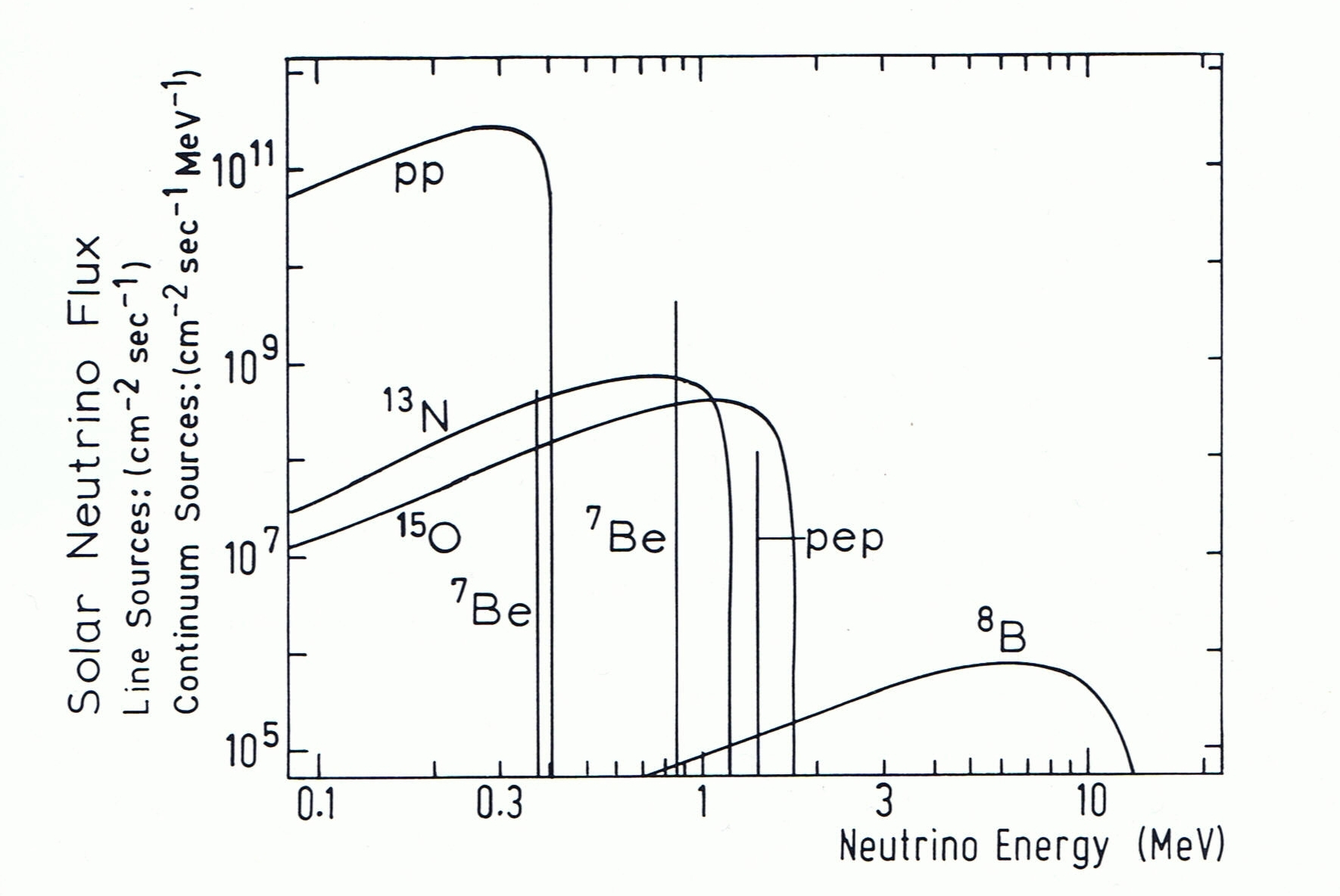}
\caption{The solar neutrino spectrum. While pp-$\nu$'s are the dominant contribution to the spectrum at low energies, the spectrum of {$^8$B}-$\nu$'s reaches a maximum energy of 15\,MeV. {$^7$Be} and pep-$\nu$'s are monoenergetic and contribute as lines to the spectrum. Also shown are the two dominant neutrino fluxes produced by the CNO cycles \cite{bah06}.}
\label{FigSolSpe}
\end{centering}
\end{figure}

Inside the pp chain, most of the neutrinos are generated in the basic fusion reaction of two protons to one deuteron (compare Fig.\,\ref{FigPpChain}). These so-called pp neutrinos appear at low energies in the solar neutrino spectrum depicted in Fig.\,\ref{FigSolSpe}, the maximum neutrino energy limited to 420\,keV. However, side branches of the pp chain allow for the production of neutrinos at higher energies. Depending on the reaction kinematics, the resulting neutrinos can either be distributed over a broad spectral range (in case of a three-particle final state), or are produced at fixed energy (if there are only two particles involved). The corresponding neutrinos are called after their production reaction; the line energies or spectral endpoint, respectively, are listed in Table\,\ref{TabNeuEne} . 

\begin{table}
\caption{Energy and predicted flux for the solar neutrinos produced along the $pp$ chain and the CNO cycle. The fluxes are derived assuming high (GS) and low (AGS) solar metallicity. Relative uncertainties are given in brackets. The last column shows the relative difference $\Delta$ in flux predictions. The fluxes are presented in units of $10^{10}$ ($pp$), $10^9$ ({$^7$Be}), $10^8$ ($pep$, {$^{13}$N}, {$^{15}$O}), $10^6$ ({$^8$B}, {$^{17}$F}), $10^3$ ($hep$) cm$^{-2}$s$^{-1}$ \cite{pen08}.}
\begin{center}
\begin{tabular}{lcccc}
\hline
Source			& Energy			& \multicolumn{3}{c}{Neutrino Flux [see caption]}  \\
				& [MeV]				& BPS08(GS)			& BPS08(AGS)		& $\Delta$ \\
\hline
pp 				& $\leq$0.420		& 5.97 (0.6\,$\%$)	& 6.04 (0.5\,$\%$)	& 1.2\,$\%$ \\
pep 			& 1.442				& 1.41 (1.1\,$\%$)	& 1.45 (1.0\,$\%$) 	& 2.8\,$\%$ \\
hep 			& $\leq$18.773		& 7.90 (15\,$\%$)		& 8.22 (15\,$\%$)		& 4.1\,$\%$ \\
{$^7$Be}		& 0.862			 & 5.07 (6\,$\%$)			& 4.57 (6\,$\%$)		& 10\,$\%$ \\
{$^8$B} 		& $\leq$14.6		& 5.94 (11\,$\%$)		& 4.72 (11\,$\%$)		& 21\,$\%$ \\
\hline
{$^{13}$N}	& $\leq1.199$	& 2.88 (15\,$\%$)				& 1.89 ($^{+14\,\%}_{-13\,\%}$) & 34\,$\%$ \\
{$^{15}$O}	& $\leq1.732$	& 2.15 ($^{+17\,\%}_{-16\,\%}$) & 1.34 ($^{+16\,\%}_{-15\,\%}$) &31\,$\%$ \\
{$^{17}$F}	& $\leq1.740$	& 5.82 ($^{+19\,\%}_{-17\,\%}$) & 3.25 ($^{+16\,\%}_{-14\,\%}$) & 44\,$\%$ \\
\hline
\end{tabular}
\label{TabNeuEne}
\end{center}
\end{table}

The competing CNO cycle fulfills the same net reaction as the pp chain, and as a consequence two neutrinos are generated per {$^4$He} fusion. There are several possible catalytic processes that are nevertheless all based on the addition of protons to a {$^{12}$C} nucleus: The corresponding neutrinos are emitted at medium energies (Fig.\,\ref{FigSolSpe}, Table\,\ref{TabNeuEne}).

The predictions for the flux contributions of different solar fusion reactions to the neutrino spectrum are based on the Standard Solar Model (SSM) \cite{bah06}. The results from spectroscopic and helioseismological observations of the Sun are used as input parameters. For a long time, astronomical data was self-consistent, while the measured electron neutrino rates were a factor 2 to 3 lower than the expected values. As will be described in Sect.\,\ref{SecNeuExp}, this inconsistency was finally resolved by the discovery of neutrino oscillations that are responsible for the observed deficit.

Currently, there are two different predictions for the observable solar neutrino flux  that are based on the deferring values for solar metallicity derived from optical spec\-tro\-sco\-py and helioseismology (Sect.\,\ref{SecSolMet}) \cite{pen08}. The fluxes are presented in Table\,\ref{TabNeuEne}, including the theoretical uncertainties. The question whether a precision measurement of solar neutrino fluxes might resolve this ambiguity will be discussed in Sect.\,\ref{SecSolMet}.

\section{Past and Present Experiments}
\label{SecNeuExp}

In 1968, the Homestake ``Chlorine'' experiment was the first to measure the neutrinos originating from solar fusion processes, an achievement for which Ray Davis was awarded the Nobel Prize in 2002 \cite{dav04}. It was the first of a series of experiments that relied on a neutrino-induced conversion of single nuclei of the element making up the target: In the case of the Homestake experiment, the reaction
\begin{eqnarray}\label{EqChlRea}
\nu_e + {^{37}\mathrm{Cl}} \rightarrow {^{37}\mathrm{Ar}} + e^- \nonumber
\end{eqnarray} 
was taking place in a target volume of 615\,t of liquid tetrachlorethylene. The minimum neutrino energy required for the conversion is 814\,keV, excluding pp-$\nu$'s from the reaction. Every month, the produced Argon was removed from the target and measured by its re-decay to Chlorine. The corresponding neutrino flux was about a factor of 3 smaller than the SSM suggested \cite{dav94}.

Two follow-up experiments, the European GALLEX experiment (later GNO) \cite{ham99,alt05} and the Russian-American SAGE experiment \cite{abd09} were based on a similar reaction on Gallium and provided an energy threshold low enough to include also the pp neutrinos in the measurement. Again, the measured rate was about half the expectation \cite{alt05}.

These radiochemical experiments had in common that they performed only an integral measurement of the solar neutrino flux, both in time and energy, as the calculated solar neutrino rate is based on counting the overall number of converted nuclei in between two extractions. Opposed to that, since the early 1980's the Japanese Kamiokande and later on Super-Kamiokande experiments were able to detect neutrinos on an event-by-event basis \cite{suz95,fuk01}: This realtime detectors used the elastic neutrino-electron scattering as detection reaction in a huge water tank. Photomultipliers mounted to the tank walls registered the Cherenkov light emitted by the recoil electrons, allowing an energy and direction-resolved measurement: The amount of light is proportional to the electron energy, while the light is emitted in a cone centered on the electron trajectory, very much like a supersonic Mach-cone. As electron and incident neutrino direction are closely correlated, the neutrino events could be clearly identified by their tracks pointing back towards the Sun. The measured rate was considerably lower than the predicted one \cite{fuk01}.

By the time the Super-Kamiokande solar data was released, there were already hints from atmospheric neutrino observation that the observed deficit in solar neutrino rates was not due to an incomplete understanding of solar neutrino production but to an intrinsic property of the particles themselves \cite{fuk98}: In the Standard Model of particle physics, every charged lepton, electron $e$, muon $\mu$, and tauon $\tau$, is accompanied by its own neutrino via weak interactions. The three kinds or ``flavors'' of neutrinos ($\nu_e$, $\nu_\mu$, $\nu_\tau$) were thought to be massless particles well separated from each other. However, the new results pointed towards the possibility that the neutrinos had the ability to periodically change flavor and to convert into each other, a phenomenon described as neutrino oscillations (e.\,g.\,\cite{kay08}).

Up to this point, all solar neutrino experiments had been exclusively or primarily sensitive to electron neutrinos. Oscillations could therefore explain the deficit in the observed event rates. The final proof was obtained in 2002, when two experiments independently confirmed the oscillation hypothesis: The Sudbury Neutrino Observatory (SNO), a water-Cherenkov detector based on heavy water, was able to measure the overall solar neutrino flux by interactions on deu\-te\-rons, independent of the neutrino flavor. The sum result corresponded to the SSM prediction \cite{ahm02}. At the same time, the liquid-scintillator detector KamLAND discovered the disappearance of electron antineutrinos emitted by the reactors of Japanese nuclear power plants: The observed energy dependence perfectly matched the predictions from oscillation theory \cite{egu03}.

\section{The Borexino Experiment}
\label{SecBorex}

With the establishment of neutrino oscillations, solar neutrino observation returns to the original objective of determining the rates at which different neutrino-producing fusion reactions occur in the solar core. The detection of solar neutrinos by the Homestake experiment proved that the Sun's energy production is fueled by thermonuclear fusion. Now, measuring the relative rates of different fusion processes by their produced neutrino flux will allow to determine solar parameters that influence the fusion rate.

A spectral measurement is mandatory in order to disentangle the contributions of different reactions to the overall neutrino flux. Of the experiments presented in Sect.\,\ref{SecNeuExp}, merely water-Cherenkov detectors are able to provide a spectral measurement of solar neutrinos. However, their sensitivity is limited to the energy range above 5\,MeV, in which only {$^8$B} and hep neutrinos contribute. Below this threshold, only the integral data of radiochemical experiments was available.

The situation changed in 2007, when the liquid-scintilla\-tor experiment Borexino came into operation \cite{ali08}. As all preceding solar neutrino detectors, it is located at an underground laboratory, in this case the Gran Sasso National Laboratories (LNGS), to shield the detector from the background of cosmic rays. The rock overburden corresponds to about 3\,500\,meters water equivalent (mwe), and only cosmic muons (and, of course, neutrinos) are able to penetrate this barrier; however, the muon flux is reduced by six orders of magnitude compared to surface level.

Like in water Cherenkov detectors, neutrinos are detected via the elastic scattering on electrons in the target volume. The recoil electrons excite the molecules of the organic scintillator solvent, which in turn de-excitate and emit light. 
The photons travel through the scintillation volume and are registered by photosensors (photomultiplier tubes, PMTs) that are mounted to the detector walls. The light output of a typical liquid scintillator is 
about 50 times greater than that of the Cherenkov effect in water, leading to both an increase in the energy resolution and a significantly lower detection threshold. This allows for the first time a spectroscopic measurement of solar neutrinos in the sub-MeV region. 

\begin{figure}[h!]
\begin{centering}
\includegraphics[width=0.4\textwidth]{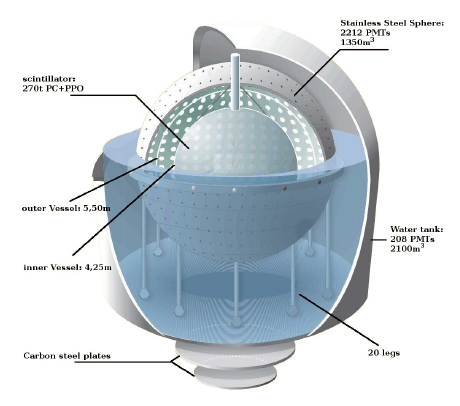}
\caption{Schematic view of the Borexino detector.}
\label{FigBorDet}
\end{centering}
\end{figure}

A major obstacle for neutrino detection is the background generated by radioisotopes dissolved in the liquid scintillator and by penetrating radiation from outside the detection volume, primarily gamma rays and cosmic muons. Therefore, the Borexino detector is structured in several concentric shielding layers of increasing radiopurity towards the central detection volume that consists of about 300 tons of liquid scintillator (Fig.\,\ref{FigBorDet}) \cite{ali08}. Purification, mainly distillation, reduced the level of radioactive impurities in the organic liquid to $10^{-18}$\,g/g in Uranium and Thorium, which is about 12 orders of magnitude less than the natural abundance. The target is contained in a 125\,\textmu m thick nylon sphere of 4.25\,m radius, the Inner Vessel. It floats in about 1000 tons of inactive organic buffer liquid which are in turn contained in a stainless steel sphere of 6.85\,m radius, to which the 2212 inward-facing PMTs are mounted. This Inner Detector is surrounded by an external water tank contained in a steel dome of 9\,m radius. This Outer Detector provides additional shielding from external radioactivity and is equipped with 208 PMTs to identify cosmic muons crossing the detector by their Cherenkov light. This active muon veto is necessary to reject muons and muon-induced events from the neutrino analysis.

\begin{figure}[h!]
\begin{centering}
\includegraphics[width=0.42\textwidth]{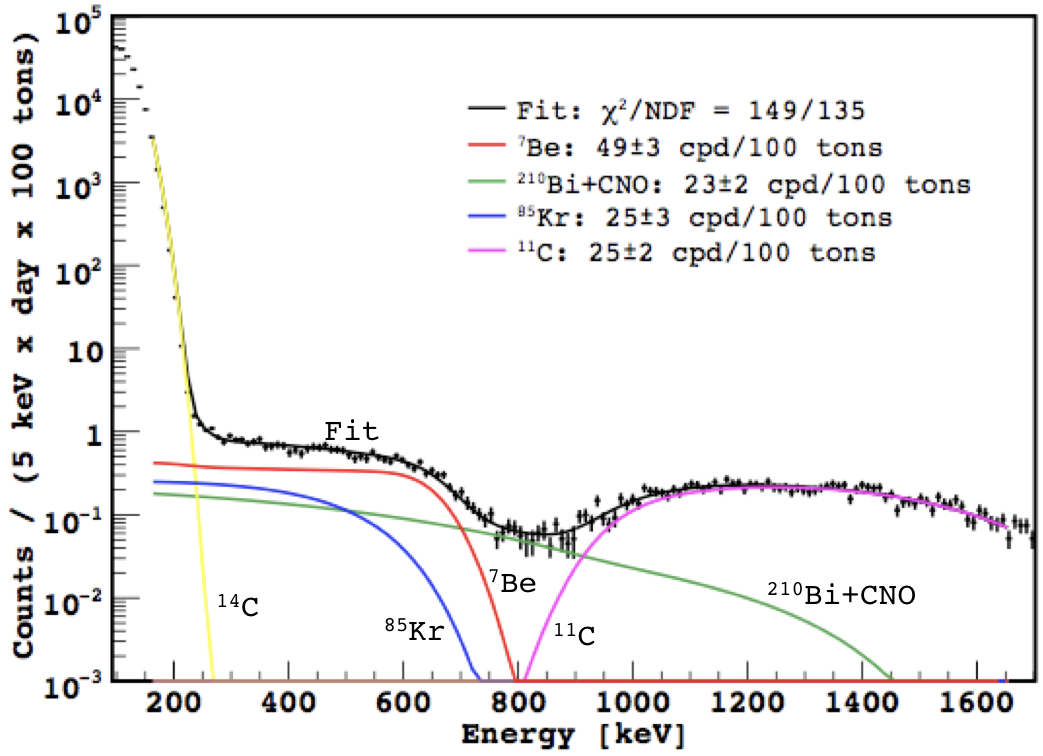}
\caption{The electron recoil spectrum below 2\,MeV based on 192 live days of Borexino data \cite{arp08b}: The black dots represent the data points, colored lines indicate the fit to the data. Both the spectra of neutrino-induced electron recoils and radioactive sources of background are included. The individual contributions are discussed in the text.}
\label{FigBe7Spe}
\end{centering}
\end{figure}

In 2008, the Borexino collaboration released data on the first flux measurement of solar {$^{7}$Be} neutrinos \cite{arp08a,arp08b}. Fig.\,\ref{FigBe7Spe} shows the electron recoil spectrum from about 0.2 to 2\,MeV, based on 192 live days of the detector. Spacial reconstruction of the events allows to restrict the analysis to the innermost 100 tons of the target; this fiducial-volume cut removes most of the background signals induced by the decay of radioactive isotopes on the vessel surface and the gamma rays of {$^{40}$K} contaminating the PMT glass of the PMTs. The information of the external veto is used to reject muon events. Coincidence signals from fast decays in the U/Th decay chains are exploited to identify further events from radioimpurities in the liquid. Finally, pulse-shape discrimination allows to remove an otherwise prominent peak caused by the $\alpha$ decay of {$^{210}$Po} at a visible energy of about 500\,keV in the spectrum.

Due to the application of these cuts, the Compton-like shoulder of electron recoil events caused by {$^{7}$Be}-$\nu$'s becomes visible at about 660\,keV in the spectrum. Superimposed are two residual sources of background: The $\beta$ decays of {$^{85}$Kr} and {$^{210}$Bi} cannot be discriminated by their pulse shape, but separated from the neutrino signal due to their known spectral shape. The {$^{7}$Be}-$\nu$ event rate obtained by a spectral fit to the data is 49$\pm$3$_{(stat)}$$\pm$4$_{(syst)}$ counts per day and 100\,tons, corresponding to about 58\,\% of the value expected from the SSM without taking neutrino oscillations into account. Correcting for the expected conversion of $\nu_e$ into $\nu_{\mu,\tau}$, the corresponding total {$^{7}$Be}-$\nu$ flux is $\Phi({^7\mathrm{Be}})=(5.18\pm0.51)\times10^9$\,cm$^{-2}$s$^{-1}$ \cite{arp08b}.

\section{Solar Neutrinos and Solar Metallicity}
\label{SecSolMet}

In recent years, a novel analysis of the solar abundances of heavy elements, i.\,e.\,the solar metallicity, has introduced inconsistencies in the astrophysical understanding of the Sun: Based on a refined modeling of the shape of the absorption lines in the solar electromagnetic spectrum, Asplund, Grevesse \& Sauval re-determined the metallicity of the solar surface to considerably lower values compared to previous analyses \cite{asp05}. Where formerly had been an excellent agreement, the new results are dramatically in conflict with helioseismological measurements. An analysis of low-degree acoustic oscillations of the Sun with BISON (Birmingham Solar-Oscillations Network) suggests that the discrepancy is not only present in the convective zone close to the solar surface but extends also to the solar core, and therefore in the production region of solar neutrinos \cite{bas07}. As fusion rates depend on the metallicity, an accurate measurement of the corresponding neutrino fluxes might be used to determine the chemical composition of the solar core region.

\begin{table}
\caption{Results of the direct measurements of solar neutrino experiments \cite{asp08b,aha08}, compared to the two SSM predictions based on different values of solar metallicity: BPS08(GS) corresponds to high metallicity, BPS08(AGS) to the novel low metallicity values \cite{pen08}. Values in brackets correspond to the relative 1$\sigma$ uncertainties. While the {$^7$Be} result seems to give a slight indication for BPS08(GS), the measured {$^8$B}-$\nu$ flux lies between the predicted values.}
\begin{center}
\begin{tabular}{lccc}
\hline
Source			& \multicolumn{3}{c}{Neutrino Flux \small{[{$^7$Be}:\,10$^9$/cm$^2$s, {$^8$B}:\,10$^6$/cm$^2$s]}}  \\
				& Experiment			& BPS08(GS)			& BPS08(AGS) \\
\hline
{$^7$Be}			& 5.18 (10\,\%)	& 5.07 (6\,$\%$)			& 4.57 (6\,$\%$)	 \\
{$^8$B} 			& 5.54 (9\,\%)	& 5.94 (11\,$\%$)		& 4.72 (11\,$\%$)	 \\
\hline
\end{tabular}
\label{TabNeuFlu}
\end{center}
\end{table}

The current status is presented in Table\,\ref{TabNeuFlu}: Up to now, merely the fluxes of {$^{7}$Be}-$\nu$'s and {$^{8}$B}-$\nu$'s were directly determined by liquid-scintillator and water Cherenkov detectors, respectively \cite{arp08b,aha08}. For comparison, the neutrino flux predictions based on the most recent SSM calculations are presented in which the solar metallicity enters as input parameter: The column BPS08(GS) denotes the fluxes corresponding to the older high values for metallicity, while BPS08(AGS) corresponds to the lower values of the new analysis \cite{pen08}. Values in brackets represent the relative uncertainties (1$\sigma$).

Neither the accuracies of the experimental results nor of the SSM predictions are for the moment sufficient to allow a definite conclusion. While the Borexino result for the {$^{7}$Be}-$\nu$ flux slightly favors the higher metallicity values, the {$^{8}$B} result lies between the two SSM predictions. However, improvement is expected on both sides: An analysis of the error budgets of the SSM neutrino fluxes reveals that especially the precision on the {$^{8}$B}-$\nu$ flux could be increased by a more accurate determination of the solar iron and carbon abundances and of the nuclear cross section $S_{1,14}$. At the same time, this would greatly decrease the uncertainty for the CNO-$\nu$ flux. The impact on the currently best-determined {$^{7}$Be}-$\nu$ flux would be low \cite{pen08}.

From the experimental side, no large improvement is expected for the {$^{8}$B}-$\nu$ flux measurements in present detectors. While the SNO experiment was terminated, Super-Ka\-mi\-o\-kan\-de results are for the moment dominated by systematical uncertainties \cite{fuk01}. Nevertheless, a new analysis of the third phase of the experiment from Aug 2006 to Aug 2008 aims on reducing the systematic effects \cite{yan09}.

The present uncertainty of the {$^{7}$Be}-$\nu$ flux as determined in Borexino is about 12\,\% \cite{arp08b}. A new analysis based on a considerably larger data set is planned for 2010. Moreover, a calibration campaign using radioactive sources placed inside the detection volume was conducted in early 2009. The results will be used to decrease the systematic uncertainties introduced by the detector response function, i.\,e.\,mainly the energy calibration, and by the position reconstruction. The latter is necessary to precisely determine the target mass used for analysis. An overall precision of better than 5\,\% on the flux measurement is envisaged.

Borexino might also contribute in a further way: In the 2008 analysis shown in Fig.\,\ref{FigBe7Spe}, the radioactive background {$^{210}$Bi} and the spectral contribution of the CNO-$\nu$'s is fitted by the same function \cite{arp08b}. This simplification is valid in the range of the {$^{7}$Be}-$\nu$ shoulder at 660\,keV where the spectral differences are small. However, the endpoint of the {$^{210}$Bi} $\beta$-decay is at 1.16\,MeV, while the spectra of some of the CNO-$\nu$'s reach up to 1.7\,MeV. Also the Compton-like shoulder caused by the monoenergetic pep-$\nu$'s should show up in this regime of the electron recoil spectrum. Unfortunately,  the background created by {$^{11}$C} $\beta^+$ decays dominates the neutrino signal between 1 and 2\,MeV by about an order of magnitude.

Currently, efforts are made to suppress this background using the signature of the production mechanism: {$^{11}$C} is a cosmogenic isotope created by cosmic muons crossing the target volume, knocking out a single neutron from a {$^{12}$C} nucleus present in the organic liquid. 
The most promising approach is a veto using both the spatial and time information of the parent muon track, the capture of the knock-out neutron on a Hydrogen nucleus that is visible in liquid-scintillator, and the decay of the {$^{11}$C} nucleus itself \cite{gal05}. At the moment, both improvements in the reconstruction software as well as in the DAQ hardware are on-going to reach the rejection efficiency of about 90 to 95\,\% that is required to resolve the spectral contributions of residual {$^{11}$C} background, CNO-$\nu$'s and pep-$\nu$'s.

\section{Potential of Future Experiments}
\label{SecFutExp}

Even if the level of cosmogenic background in Borexino proves to be too large for an actual measurement of the CNO and pep neutrino fluxes, the best limits on their spectral contributions will significantly be improved. However, the currently upcoming SNO+ experiment will be in a much better position to measure these contributions: Based on the setup of the terminated SNO experiment, the acrylic vessel formerly holding the heavy water target will be refilled with liquid scintillator. The resulting target mass will be three times larger than in Borexino. More importantly, SNO+ will be located at a depth corresponding to 6000\,mwe in the Creighton Mine at Sudbury, the additional shielding reducing the residual muon flux by about 2 orders of magnitude compared to the Borexino site. Therefore, the background of cosmogenic {$^{11}$C} will play a minor role in SNO+, the achievable accuracy mainly depending on the radiopurity of the liquid scintillator and the detector materials.

\begin{figure}[h!]
\begin{centering}
\includegraphics[width=0.39\textwidth]{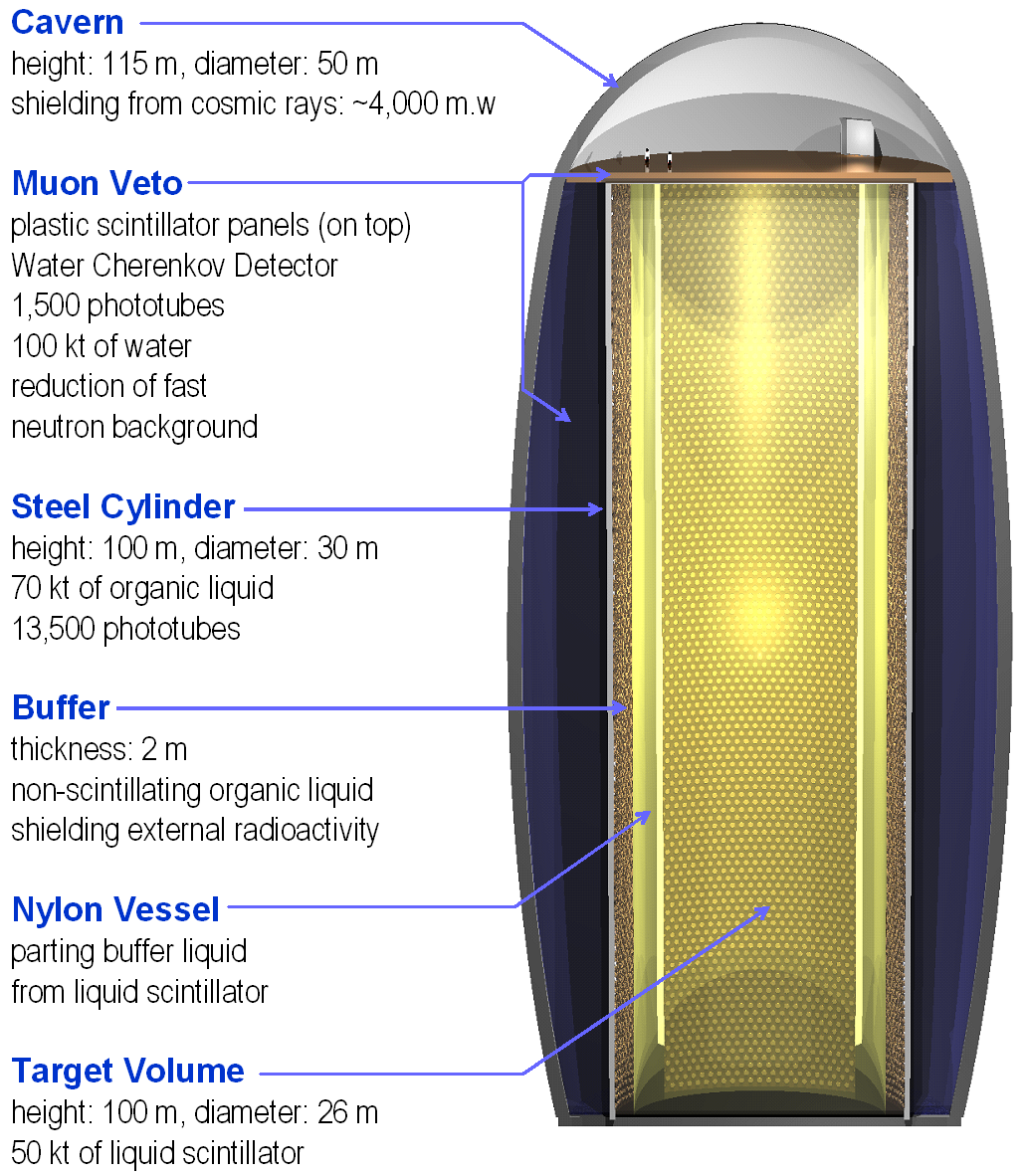}
\caption{Envisaged layout of the LENA detector.}
\label{FigLenDet}
\end{centering}
\end{figure}

Further in the future, the LENA (Low Energy Neutrino Astronomy) experiment is planned to go into operation in 2020 \cite{mar08}. A schematic of the setup is shown in Fig.\,\ref{FigLenDet}. With a target of 50 kilotons of liquid scintillator, LENA will be an observatory for a variety of terrestrial and astrophysical neutrino sources: About 15000 neutrino events are expected in case of a core-collapse Supernova (SN) at the center of our galaxy, giving insights into the nuclear processes leading to the collapse of the stellar core region and the subsequent neutrino cooling of the proto-neutron star \cite{wur07p}. Even in the absence of a galactic SN, the faint neutrino flux generated by extra-galactic SNe offers a possibility to obtain information both on the star formation rate and the SN neutrino spectrum. While the event rates caused by this diffuse SN neutrino background are too low to be discovered in present-day detectors, the sensitivity of LENA will be sufficient to identify about 10 events per year \cite{wur07}. Also the chemical composition and heat production of our own planet can be investigated by a precision measurement of geoneutrinos \cite{hoc07}: These neutrinos are produced by the radioactive elements of the natural Uranium and Thorium decay chains that are embedded in the Earth's crust and mantle.

Regarding the spectroscopy of solar neutrinos, a new era of precision measurements will begin: Based on a fiducial volume of 20\,kt, rates of several hundred events per day are expected for CNO, pep and {$^8$B} $\nu$'s in LENA, allowing a precise determination of their relative contributions to the solar neutrino spectrum \cite{wur07p}. As LENA will be placed at greater depth than Borexino (albeit not as deep as SNO+), cosmogenic background levels will be significantly lower. Most remarkable is probably the daily rate of 10\,000 {$^7$Be}-$\nu$ events per day: These statistics would allow to search for modulations in the neutrino flux on a per mill level. Such modulations could be induced by the interaction of the propagating neutrinos with solar or terrestrial matter \cite{ble05}, or by periodical changes in the neutrino production rate itself. The range of interest reaches from periods of tens of minutes typical for helioseismic g-modes to the scale of decades correlated to the solar cycle \cite{kra90}.
~\\
\section{Conclusions}

In the last 20 years, the spectroscopy of solar neutrinos has greatly enhanced our knowledge of neutrino particle parameters, but has also confirmed the elaborate models describing the solar energy production by fusion in the pp-chain and the CNO cycle. While the currently available data on solar neutrino fluxes is not accurate enough to resolve the conflict in the determination of solar metallicity, a considerable improvement in the near future can be expected. 

Up to now, only {$^7$Be} and {$^8$B} neutrino fluxes have spectroscopically been measured. However, there is a good chance that the contributions of pep and CNO neutrinos will be discovered either in the currently running Borexino detector or in the upcoming SNO+ experiment. In 10 years from now, a large liquid-scintillator based neutrino observatory like LENA might allow precision measurements of the solar neutrino spectrum, possibly revealing time variations in the solar energy production. For sure, the next years will deepen both our understanding of neutrinos as well as of their sources.
~\\
\acknowledgements
This work was supported by the Deutsche For\-schungs\-ge\-sell\-schaft, the Munich Cluster of Excellence ``Universe'', and the Maier-Leibnitz-Laboratorium in Garching. We would like to thank the Borexino collaboration for the common work.


\end{document}